\documentstyle[psfig]{l-aa}

\newcommand{\degrees}{$^{\circ}$}

\newcommand{\OI}{\mbox{O\,{\sc i}}}
\newcommand{\NeII}{\mbox{Ne\,{\sc ii}}}
\newcommand{\FeII}{\mbox{Fe\,{\sc ii}}}

\def\mum{\hbox{\,$\mu$m}}

\def\fluxi{\hbox{\,erg\,\,s$^{-1}$\,cm$^{-2}$}}
\def\flux{\hbox{\,erg\,\,s$^{-1}$\,cm$^{-2}$\,sr$^{-1}$}}
\def\fluxpm{\hbox{\,erg\,\,s$^{-1}$\,cm$^{-2}$\,\mum$^{-1}$\,sr$^{-1}$}}

\newcommand{\srce}{\mbox{IC\,{\sc 443}}}

\begin{document}
\thesaurus{ }

\title {ISOCAM spectro-imaging of the H$_2$ rotational lines in the supernova 
        remnant \srce~ \thanks{Based on observations with ISO, an ESA project 
        with instruments funded by ESA Member States (especially the PI 
        countries: France, Germany, the Netherlands and the United Kingdom) 
        and with the participation of ISAS and NASA.} 
}

\author{D. Cesarsky\inst{1}, P. Cox\inst{1}, G. Pineau des For\^ets\inst{2},
	 E.F. van Dishoeck\inst{3}, F. Boulanger\inst{1}, C.M. Wright\inst{4}}

\offprints{P. Cox (cox@ias.fr)}

\institute{$^1$Institut d'Astrophysique Spatiale, B\^{a}t. 120,
           Universit\'e de Paris XI, 91405 Orsay, France \\
	   $^2$DAEC, Observatoire de Paris, F-92195 Meudon Principal Cedex, France \\
	   $^3$Leiden Observatory, P.O. Box 9513, 2300 RA Leiden, The Netherlands \\
	   $^4$School of Physics, University College, ADFA, UNSW, Canberra ACT 2600, Australia
}

\date{Received April 1999; accepted  Hopefully in 1999}

\maketitle

\markboth {D. Cesarsky et al.~ H$_2$ rotational lines in \srce}
          {D. Cesarsky et al.~ H$_2$ rotational lines in \srce}
\begin{abstract}

We report spectro-imaging observations of the bright western ridge of the
supernova remnant IC~443 obtained with the ISOCAM circular variable filter
(CVF) on board the {\em Infrared Space Observatory (ISO)}. This ridge
corresponds to a location where the interaction between the blast wave of the
supernova and ambient molecular gas is amongst the strongest. The CVF data
show that the 5 to 14\mum~ spectrum is dominated by the pure rotational lines
of molecular hydrogen (v = 0--0, S(2) to S(8) transitions).
At all positions along the ridge, the H$_2$ rotational lines are very strong
with  typical line fluxes of  $\rm 10^{-4} \, to \,  10^{-3}$\,\flux.
We compare the data to a new time-dependent shock model; the 
rotational line fluxes in IC~443 are reproduced 
within factors of 2 for evolutionary times between 1,000 and 2,000 years with a
shock velocity of $\rm \sim 30 \, km \, s^{-1}$ and a pre-shock density
of $\rm \sim 10^4 \, cm^{-3}$.


\end{abstract}

\section{Introduction}

The supernova remnant IC~443 is a prime example of the interaction of a
supernova blast wave with an ambient molecular cloud. On optical plates, \srce~
appears as  an incomplete shell of filaments (Fig.~1) with a total extent of
about 20~arcmin, i.e. $\rm \sim 9\, pc$ for an adopted distance of 1500\,pc.
The shock generated by the  supernova explosion, that occurred  $\rm (4-13)
\times 10^3 \, years$ ago, encountered nearby  molecular gas which is mainly
found along a NW-SE direction across the face of the optical shell. \srce~has
been the subject of numerous studies from X-rays, visible, infrared to radio
wavelengths (e.g., Mufson et al., 1986 and references therein). Studies
of the interaction  between the shock and the ambient molecular gas were done
by observing molecular hydrogen in the rotational--vibrational transitions 
(Burton 1988, Burton et al. 1990- see Fig.\,1 - and
Richter et al. 1995a), in the pure rotational S(2) transition (Richter et al.
1995b),   in other simple molecules such as CO and HCO$^+$ (e.g., van Dishoeck
et al. 1993 and references therein)  and in atomic carbon (Keene at
al. 1996).

In this letter we report mapping results of the pure rotational lines of H$_2$ 
using the ISOCAM CVF over the western ridge of IC~443, a position corresponding
to clump G in the nomenclature of Huang et al. (1986). The observations reveal 
the details of the structure and  the physical conditions of the shocked
molecular gas in IC~443 with a pixel field of view of $\rm 6^{\prime\prime}$
and at unprecedented sensitivity (mJy). The observed H$_2$ line fluxes are well
predicted by a time-dependent shock model recently developed by Chi\`eze et al.
(1998) and Flower \& Pineau des For\^ets (1999).

\section{Observations and data reduction}

The observations were done in 1998 February with the ISOCAM CVF (Cesarsky et
al. 1996). A pixel size of $\rm 6^{\prime\prime}$ was used, yielding a total
field of view of  $\rm 3^{\prime} \times 3^{\prime}$. Scans of the
long-wavelength CVF were obtained: the LW-CVF2 from 13.53 to 9\,\mum~ and
then the LW-CVF1 from 9 to 5.0\,\mum~ for a total of 115 wavelength steps;
the resolving power is 40.
Each wavelength was observed for 12.6 sec, i.e. six readouts at 2.1 sec, for a
total observing time of some 30 minutes.

ISOCAM was pointed towards $\alpha$ = 06h~13m~41.0s and $\delta$ =
22\degrees~33\arcmin~10.4\arcsec~(coordinates B1950.0), a position
corresponding  to the center of the molecular clump G which is also a peak in
the H$_2$ emission (Fig.\,1).

\begin{figure}[!ht]
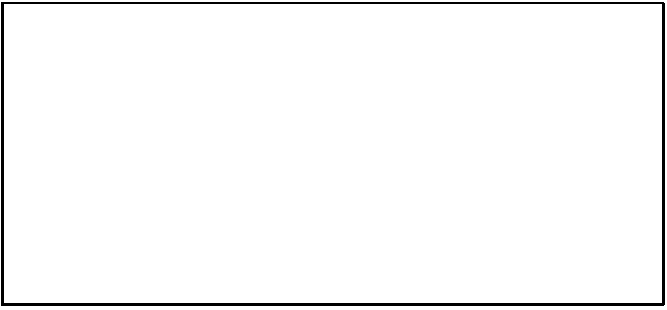

\picplace{4cm}
 \caption{The footprint of ISOCAM (box) depicted against the DSS image of the IC\,443 supernova 
	  remnant together with the H$_2$~1--0S(1) emission (contours) from Burton et al. (1990).}
\end{figure}

The data reduction includes a new time dependent dark current correction
(Biviano \& Sauvage, in preparation) as well as a new correction procedure for
the transients (Coulais \& Abergel 1999). The photometric calibration was done
using the calibration files applicable to the current release of the ISOCAM
off-line processing software (V7.0); the absolute calibration is conservatively
estimated to be on the order of 25\%.

The zodiacal background was subtracted  adopting the spectrum obtained by Reach
et al. (1996; but note that the published data were reduced again using the
same algorithms  and calibration files as applied to the current data) with a
scaling factor of 0.7.

The flux of the H$_2$ lines was obtained by numerical integration under the
line profile, after subtracting a linear baseline defined by several points at
either side of the line profile; the line fluxes thus obtained are independent
of the assumed zodiacal background. We estimate the statistic uncertainty of
the numerical  integration by applying the same integration algorithm to
regions of the  spectrum devoid of any spectral emission; we consistently obtain
an rms noise of some 2\,10$^{-5}$ \flux. 

\section{Results}

The 5 to 14\,\mum~ spectrum of the molecular clump G in \srce~ is dominated by
the series of the pure ($v=0-0$)  rotational lines of molecular hydrogen from
the S(2) to the S(8) transitions (Fig.~2). In particular, there is no
indication of any  atomic fine structure line. At low level intensity, the
set of dust emission bands from 6.2 to 11.3\,\mum~ is clearly present,
including the 12.7 \mum~ band. Note that the relative strength of the dust
bands is typical of that of the ISM (Boulanger 1998), suggesting that a
possible contribution from the [\NeII] line at 12.8 \mum~ is negligeable. This
dust emission has comparable intensities over the entire ISOCAM field, i.e. a
few MJy\,sr$^{-1}$, which is similar to the intensities measured in  diffuse
regions of the Ophiucus cloud. An image in one of the dust bands, e.g.
7.7~$\mu$m, does not show any structure correlated with the molecular filament. 
Using the trend between the  band intensity and the UV radiation field shown by
Boulanger (1998), we  find that the excitation of the dust bands is
commensurable with a  radiation field $\rm G \sim 10 \times G_0$.  The dust
bands are probably unrelated to the supernova remnant and more likely mixed 
with the interstellar gas along the 1.5~kpc line of sight toward  \srce.
Although very faint, the 7.7 and 6.2\,\mum~ dust bands  contaminate the S(4)
and S(6) H$_2$ lines. We have  fitted Lorentzians to  the dust bands (Boulanger
et al. 1998) and removed the resulting mean dust band  spectrum from each
individual CVF spectrum prior to performing the numerical integrations.
\begin{figure}[!h]
	\begin{center}
\psfig{figure={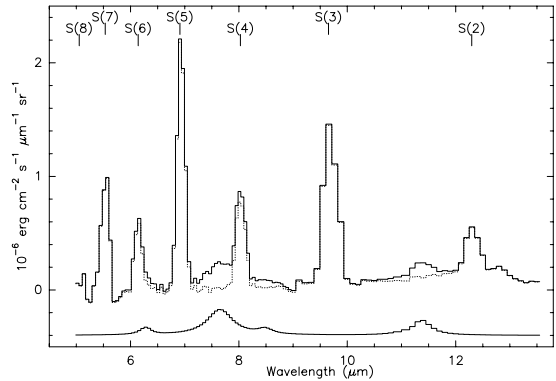},width=8cm,angle=-90} 
	\end{center}
\vspace{-1.5cm}
\caption {IC443 mean CVF spectrum, i.e. integrated over the observed field, 
before (solid line) and after (dotted line) subtraction of the dust bands
spectrum which is shown at the bottom of the figure at true  scale but shifted
down by 4\,10$^{-7}$\,\fluxpm. The rotational lines of molecular hydrogen are
labeled.}
\end{figure}

\begin{figure*}
\picplace{4cm}
\caption{The distribution of the emission of the S(2) to S(7) H$_2$ lines
towards Clump G in \srce~ (upper left panel). The next top panels show the
emission in the S(2) and S(7) H$_2$ lines.  Contours are drawn at the
10\%, 20\%,  etc. level. The corresponding peak strengths are  8.1~10$^{-3}$,
7.9~10$^{-4}$ and 2.2~10$^{-3}$\,\flux~ from left to right. The middle panels
show the ISOCAM CVF   spectra towards the three emission peaks A, B, and C. The
lower panels present the excitation diagrams after correcting for an extinction
of $\rm A_{2.12 \, \mum} = 0.6 \, mag$. Note that the spectra correspond to one
pixel whereas the excitation diagrams have been built from a $3 \times 3$ pixel
box car smoothed data. Each excitation diagram has been fitted with a 
two-component model involving `warm' (dashed line) and `hot' (dotted line)
H$_2$.  The full lines represent the sum of both H$_2$ components - see text
and Table~1 for details.} 
\end{figure*}

Figure~3 shows the total emission of the H$_2$ lines between 
5 and 13.5\,$\mu$m, i.e. the sum of the S(2) to S(7) lines, together
with the integrated line intensity of the S(2) and S(7) transitions (top panels).
The H$_2$ emission is found along a ridge of about
$30^{\prime\prime}\times80^{\prime\prime}$ (0.25\,pc$\times$0.65\,pc)  running
SW to NE, a structure which is comparable  to that seen in CO or HCO$^+$ (van
Dishoeck et al. 1993, Tauber et al. 1994).  The higher spatial resolution of
the ISOCAM data clearly reveal a series of  
knots sitting on a plateau.  The H$_2$ knots are very bright with
peak values of a few $10^{-3}$~\flux~ (Table~1). The eastern side of the
molecular ridge, facing the origin of the supernova explosion, appears sharper
than the opposite side where weak emission is found extending westwards.

\section{Discussion}

Altogether there are about 130 pixels that show H$_2$ emission with intensities
above the 10$\, \sigma$ level, i.e.  $\rm > 2\, 10^{-4}$ \, \flux~ for all the
six rotational transitions S(2) to S(7). This fact allows the construction of
as many H$_2$ excitation diagrams which plot   the logarithm of the column
density, corrected for statistical weight,  in the upper level of each H$_2$
transition vs. the energy of that level, $\rm E_u$. When calculating the column
densities, we corrected the observed line fluxes for extinction, using a screen
model and the extinction curve from Draine \& Lee (1984).  Peak\,A corresponds
to a position studied by Moorhouse et al. (1991) and Richter et al. (1995a -
their Position 3). Moorhouse et al. derived an  extinction of  A$_{2.12 \,
\mum}$ = 0.6 \, mag., whereas Richter et al. estimated a somewhat higher value
($\sim \,1$\,mag.).  We adopted the former value (see dereddening factors in
Table~1). The middle and bottom panels in Fig.~3 present the CVF spectra and 
the corresponding excitation diagrams of the three emission peaks labeled A, B,
C. The statistical weights used in Fig.~3  include a factor of 3 for
ortho-H$_2$ and 1 for para-H$_2$.

\begin{table*}		
\caption{Line fluxes and derived parameters from the model fits - see text}
\begin{center}
\leavevmode
\begin{tabular}[h]{cccccccccccc}
\hline
\hline
\multicolumn{12}{c}{} \\
\multicolumn{1}{l}{H$_2$ Transition} &
\multicolumn{1}{c}{S(2)} &  \multicolumn{1}{c}{S(3)} &
\multicolumn{1}{c}{S(4)} & \multicolumn{1}{c}{S(5)}  & 
\multicolumn{1}{c}{S(6)} & \multicolumn{1}{c}{S(7)}  &
\multicolumn{1}{c}{~}    & \multicolumn{4}{c}{Model fit} \\
\multicolumn{1}{l}{Wavelength [\mum]} &
\multicolumn{1}{c}{12.28}          &  \multicolumn{1}{c}{9.66} &
\multicolumn{1}{c}{8.02}           &  \multicolumn{1}{c}{6.91} & 
\multicolumn{1}{c}{6.11}           &  \multicolumn{1}{c}{5.51} &
\multicolumn{1}{}{} &              \multicolumn{2}{c}{Warm Component} & 
\multicolumn{2}{c}{Hot Component} \\
\multicolumn{1}{l}{Dereddening factor$^1$} &
\multicolumn{1}{c}{1.16}   &  \multicolumn{1}{c}{1.42} &
\multicolumn{1}{c}{1.13}   &  \multicolumn{1}{c}{1.07} & 
\multicolumn{1}{c}{1.08}   &  \multicolumn{1}{c}{1.09} &
\multicolumn{1}{}{}        &  \multicolumn{1}{c}{T}    &  
\multicolumn{1}{c}{N$\rm_{H_2}$} &
\multicolumn{1}{c}{T}      &  \multicolumn{1}{c}{N$\rm_{H_2}$} \\
\multicolumn{1}{l}{Peak identification} & 
\multicolumn{6}{c}{Observed line flux [$10^{-4}$\,\flux]}  &  
\multicolumn{1}{}{}        &  \multicolumn{1}{c}{[K]}      &  
\multicolumn{1}{c}{[10$^{20}$ cm$^{-2}$]}                  &
\multicolumn{1}{c}{[K]}    &  \multicolumn{1}{c}{[10$^{20}$ cm$^{-2}$]} \\
\cline{1-7}  
\cline{1-7}
\cline{9-12} 
\cline{9-12}
\multicolumn{12}{c}{} \\
A & 3.7 & 14.3 & 7.7 & 17.6 & 4.4 &  8.3 & & 657 & 2.2   & 1288 & 0.22 \\
B & 5.1 & 11.9 & 5.3 & 18.9 & 5.1 & 10.3 & & 330 & 19.7  & 1172 & 0.48 \\
C & 2.4 &  8.4 & 4.5 & 12.4 & 3.3 &  6.2 & & 446 & 2.6   & 1115 & 0.37 \\
\multicolumn{12}{c}{} \\
\hline
\end{tabular}
\end{center}
\footnotesize{\hspace{8mm} $^1$ $\rm 10^{0.4 \, A_{\lambda}}$ adopting
the extinction curve of Draine \& Lee (1984) and 
$\rm A_{2.12 \, \mum} = 0.6 \, mag.$}
\end{table*}

The excitation diagrams for Peaks A, B, and C show that a single excitation
temperature does not reproduce the H$_2$ lines observed in \srce~ and that
emission from gas with a range of temperatures is required. The results of a
simple LTE two-component H$_2$ model are shown in Fig.~3: a `warm' H$_2$ component
with an excitation temperature of $\rm \sim 500 \, K$ and typical column
densities  in between $10^{20}$ and $\rm 10^{21} cm^{-2}$, and a `hot' H$_2$ 
component with $\rm T_{ex} \, \sim \, 1200 \, K$ and  $\rm N_{H_2}$ a few $\rm
10^{19} \, cm^{-2}$ (Table~1). The parameters of the `warm' component are
determined almost entirely by the intensities of the S(2) and S(3) lines and 
the `hot' component dominates the H$_2$ transitions S(4) to S(7).  Although the
evidence for a `warm' component is very strong, the ISOCAM data only poorly
constrain its properties because of the lack of measurement of the S(1) and
S(0) H$_2$ transitions. Furthermore, the uncertainty in the extinction
correction (especially for the S(3) line whose position coincides with the peak
of the silicate 9.7\, \mum~band) introduces an additional uncertainty in the
temperature determination. Using different extinction laws and adopting values
for A$_{2.12\,\mum}$ between 0.5 and 1\,mag.,  we derive typical
uncertainties of $\rm \pm \, 100$ and $\rm \pm \, 250 \, K$ for the warm and
the hot components, respectively. 

\begin{figure}
\centerline{\psfig{figure={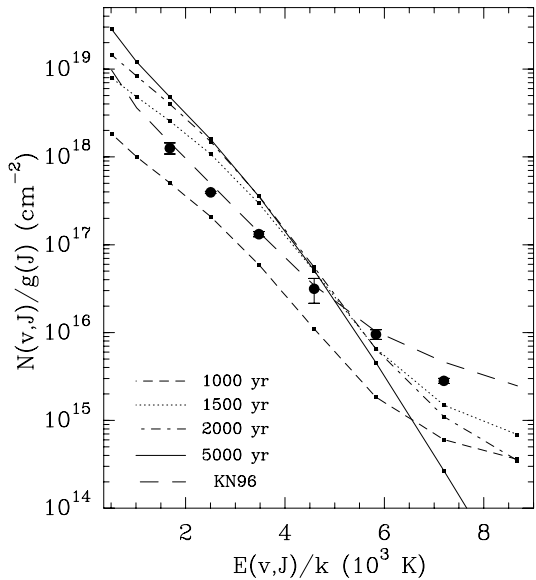},width=10cm}}
\vspace{-2cm}
\caption{The excitation diagram for Peak A (filled circles)
compared to the predictions of the time-dependent shock model 
of Chi\`eze et al. (1998) with a pre-shock density of $\rm 10^4 \, cm^{-3}$,
a shock velocity of $\rm 30 \, km \, s^{-1}$ and four evolutionary times.
The long-dashed curve labelled (KN96) presents the predictions of
a model with two C-shocks from Kaufman \& Neufeld (1996) - see text.} 
\end{figure}

ISO spectroscopy of regions where shocks dominate the excitation has revealed
that the shocked gas has a range of temperatures from a few 100\,K to several
1000\,K: Cepheus\,A (Wright et al. 1996),  Orion (Rosenthal et al. 1999) and
bipolar outflows  (Cabrit et al. 1999). The H$_2$ lines cannot be explained by
a single  shock model and combinations of J- and/or C-type shocks with
different  velocities and pre-shock densities have been invoked to account for 
the observed line fluxes. Similar conclusions have been reached for  \srce.
Based on an analysis of the [\OI] 63\,\mum~ fine structure line and
near-infrared H$_2$ lines, Burton et al. (1990) concluded that the infrared
line emission of  \srce~ can only be modeled as a  slow (10-20$\rm \, km \,
s^{-1}$), partially dissociating J shock where the oxygen chemistry is
suppressed, i.e. the cooling is dominated by [\OI] emission and not by H$_2$O
cooling - see also Richter et al. (1995a, b). Far-infrared spectroscopy
obtained with ISO on IC\,443 confirms these conclusions and will be discussed
in a forthcoming paper.

Following the interpretation of Wright et al. (1996) for the shocked H$_2$ gas
in Cepheus\,A, the H$_2$ lines in  \srce~ can be fit by a combination of 
two C-shocks from the models of Kaufman \& Neufeld (1996). A good match to the 
data towards Peak\,A is obtained combining a first shock with a pre-shock density
of $\rm 10^4 \, cm^{-3}$, a velocity of $\rm 20 \, km s^{-1}$ and a covering factor
$\Phi$ of 0.85 with a second shock of  $\rm 10^6 \, cm^{-3}$, $\rm 35 \, km s^{-1}$ 
and $\Phi$ of 0.008 (see Fig.~4). Such a steady-state model requires at least 
two C-shocks with a set of 3 free parameters and relatively high pre-shock 
densities for the high velocity component.  

Recently, Chi\`eze et al. (1998) pointed out that the intensities of the
ro-vibrational H$_2$ lines are sensitive to the temporal evolution of a shock
wave.  In many astrophysical situations, shock waves are unlikely to have
reached steady-state which occurs at approximately 10$^4$\,yr.  At times scales
of a few 10$^3$\,yr, the shocked gas may show both  C- and J-type
characteristics: within the C-shock, a J-type shock is  established heating a
small fraction of the gas to high temperatures  (Flower \& Pineau des For\^ets
1999). In the case of  \srce, the typical  size of the molecular clump G is
$\leq 20^{\prime\prime}$, i.e.  $\rm \leq 2 \times 10^{17} \, cm$, but some of the
molecular clumps have  typical sizes of about a few arcsecs (Richter et al.
1995a).  For a shock velocity of $\rm 20-30 \, km s^{-1}$, the crossing time of
the shock wave is thus $\rm \leq 2500 - 4000 \, yr$  indicating that the shock
wave in  clump G  is not in steady-state.

Figure~4 shows the predictions of the time-dependent shock model for an
observation along the direction of the shock propagation. The model results are
given for four evolutionary times with the following parameters: pre-shock gas
density $\rm n_{H} = 10^{4} \, cm^{-3}$, shock velocity $\rm v_{s} = 30 \, km
s^{-1}$, magnetic field strength $\rm B = 100 \, \mu G$ and ortho-to-para ratio
of 3. The filling factor in this model is equal to 1. A smaller filling factor
could be compensated by a larger line of sight path across the shocked H$_2$
gas for a non face-on shock.  The post-shock densities are $\rm 10^5 - 10^6 \,
cm^{-3}$ comparable with the values derived by van Dishoeck et al. (1993). The
agreement between the model predictions and the observations is best for
intermediate times, i.e. $\le$\,2000\,yr. At earlier epochs when the J-shock
dominates, the low excitation H$_2$ lines are too weak. And after 5000\,yr,
when the C-shock steady-state is reached, the higher excitation H$_2$ lines
(above 10$^4$\,K) are much too weak. In between, the intensities of the H$_2$
rotational lines  are predicted within factors of 2 and the coexistence of the
`hot' and `warm' H$_2$  components is well explained within a single model. 
Models with other parameters (e.g., $\rm n_H = 3 \times 10^3 \, cm^{-3}$, $\rm
v_s = 35 \, km s^{-1}$) provide less  good fits. The best fits are obtained for
early evolutionary times ($\sim$\,1000-2000\,years) and densities of $\rm \sim
10^4 \, cm^{-3}$ with shock velocities $\rm v_s = 30 - 40 \, km s^{-1}$. Higher
shock velocities will predict too large intensities for the high-excitation
H$_2$ lines. 

The agreement between these time-dependent model predictions and the data of
\srce~is very encouraging in view of some of the simplistic underlying assumptions.
In particular, the assumed geometry (plane parallel) is oversimplified and does
not describe the molecular filament which is seen edge-on and consists of 
numerous small clumps. Clearly a more thorough study should be done to
explore the entire parameter space of the model and compare the predictions 
with additional data available on \srce.

In the model, the S(2) to S(7) lines account for about 70\% of the luminosity
in all the H$_2$ lines. At peak A, the measured S(2) to S(7) flux is $\rm \sim
4.6 \times 10^{-12}$ \,\fluxi\,  ($\rm \sim 0.4 \, L_{\odot}$) and, according
to the model, the H$_2$ lines alone would thus carry $\rm \sim 0.6 L_{\odot}$. 
Towards clump G, the S(3) and S(2) H$_2$ lines account for almost the entire
IRAS 12\,\mum-band emission. The mean value of these lines in that band  is
5.4\,$\rm MJy \, sr^{-1}$ comparable to the IRAS peak value, i.e. 6\,$\rm MJy
\, sr^{-1}$ (e.g.,  Oliva et al. 1999).  Similarly, the IRAS 25\, \mum-band
could also be due to H$_2$ line emission. The model predictions (Fig.~4) for
the S(0) and S(1) line fluxes (at an evolutionary time of 2000 years) are
1.2\,$10^{-5}$ and  7.7\,$\rm 10^{-4}$ \, \flux, respectively. Taking into
account the 20\%  transmission  at 17.03\,\mum~of the IRAS 25\, \mum~band, the
strong S(1) line would thus contribute $\rm \sim 4 \, MJy \, sr^{-1}$  at
25\,\mum, which is comparable to  the measured 25\,\mum~IRAS flux at Peak\,A
($\rm \sim \, 4.5 \, MJy \, sr^{-1}$, e.g. Oliva et al. 1999). These results
strongly suggest that the excitation of the gas in \srce~ is entirely
collisional. 

Finally, Oliva et al. (1999) found towards the optical filaments of \srce~,
which trace the low density atomic gas, that most of the 12 and 25\,\mum~ IRAS
fluxes is  accounted for by ionized line emission (mainly [\NeII] and [\FeII]).
Our results show that this conclusion cannot be generalised towards  the
molecular hydrogen ring (Fig.~1) where the dense molecular gas  essentially
cools via the H$_2$ lines in the   near- and mid-infrared and via the [\OI]
emission line   in the far-infrared.

{\it Acknowledgments:}  Michael Burton is kindly thanked for providing his
H$_2$ map of \srce.

\end{document}